\documentclass[prb,10pt,twocolumn]{revtex4}

\usepackage{graphicx}

\begin{document}

\title{Measurements of the Carrier Dynamics and Terahertz Response of Oriented Germanium Nanowires using Optical-Pump Terahertz-Probe Spectroscopy}

\author{Jared H. Strait}
\affiliation{School of Electrical and Computer Engineering, Cornell University, Ithaca, NY 14853}
\email{jhs295@cornell.edu}
\author{Paul A. George}
\affiliation{School of Electrical and Computer Engineering, Cornell University, Ithaca, NY 14853}
\author{Mark Levendorf, Martin Blood-Forsythe}
\affiliation{Department of Chemistry and Chemical Biology, Cornell University, Ithaca, NY 14853}
\author{Farhan Rana}
\affiliation{School of Electrical and Computer Engineering, Cornell University, Ithaca, NY 14853}
\author{Jiwoong Park}
\affiliation{Department of Chemistry and Chemical Biology, Cornell University, Ithaca, NY 14853}

\begin{abstract}
We have measured the terahertz response of oriented germanium nanowires using ultrafast optical-pump terahertz-probe spectroscopy. We present results on the time, frequency, and polarization dependence of the terahertz response. Our results indicate intraband energy relaxation times of photoexcited carriers in the 1.5-2.0 ps range, carrier density dependent interband electron-hole recombination times in the 75-125 ps range, and carrier momentum scattering rates in the 60-90 fs range. Additionally, the terahertz response of the nanowires is strongly polarization dependent despite the subwavelength dimensions of the nanowires. The differential terahertz transmission is found to be large when the field is polarized parallel to the nanowires and very small when the field is polarized perpendicular to the nanowires. This polarization dependence of the terahertz response can be explained in terms of the induced depolarization fields and the resulting magnitudes of the surface plasmon frequencies.
\end{abstract}

\maketitle

\section{Introduction}
In recent years, semiconductor nanowires have gathered much interest. Nanowires have been applied to an array of applications that highlight their versatility as building blocks of integrated electronics (transistors) and photonics (waveguides, lasers, photodetectors, solar cells) \cite{Lieber09,Lieber08,Sirbuly05,Ahn07,Law04,Agarwal06}. Germanium nanowires are of particular interest due to the attractive material properties of Germanium, including large electron and hole mobilities and large optical absorption in the visible/near-IR.  These properties could make germanium nanowires the choice for next generation electrical and photonic devices, such as transistors, CMOS compatible photodetectors, and solar cells.  Understanding the fast electrical and optical response as well as ultrafast dynamics of carriers in nanowires is important for most of the applications mentioned here. In this paper, we present results on the measurement of the terahertz (THz) response as well as ultrafast carrier dynamics in photoexcited germanium nanowires using optical-pump THz-probe spectroscopy.  

Ultrafast carrier dynamics in group III-V, II-VI, and group IV semiconductor nanowires have been studied with optical-pump optical-probe spectroscopy measurements \cite{Prasankumar08,Sun05} which are sensitive primarily to the carrier occupation of specific regions in the energy bands. Optical-pump THz-probe spectroscopy, in which the probe photon energy is $\sim$5 meV, is sensitive to not only the total carrier density but also to the distribution of these carriers in energy within the bands.  The latter is true since the energy distribution of carriers affects the THz optical conductivity \cite{Paul08}. Optical-pump THz-probe spectroscopy can therefore be used to simultaneously study both intraband relaxation and interband recombination dynamics of photoexcited electrons and holes on ultrafast time scales. Our results show intraband carrier relaxation rates (attributed to intravalley and intervalley phonon scattering) in the 1.5-2 ps range and carrier density-dependent recombination rates (attributed to nanowire surface defects) in the 75-125 ps range at room temperature in 80 nm diameter wires.  

The fast electrical response of nanowires at THz frequencies can also be studied with optical-pump THz-probe spectroscopy \cite{Parkinson07}. With this technique, we measure carrier momentum scattering times in the 60-90 fs range.  Additionally, we find the THz response of oriented nanowires to be strongly dependent on the polarization of the THz field. The differential THz transmission through photoexcited nanowires is most affected when the THz field is polarized parallel to the nanowires, while no appreciable response is detected when the THz field is polarized perpendicular to the nanowires. The shape anisotropy of the nanowires at subwavelength scales leads to a strong polarization dependent macroscopic THz response. Our results indicate the possibility of realizing optically or electrically controlled active THz devices based on semiconductor nanowires.

\begin{figure}[tbp]
	\centering
		\includegraphics[width=.40\textwidth]{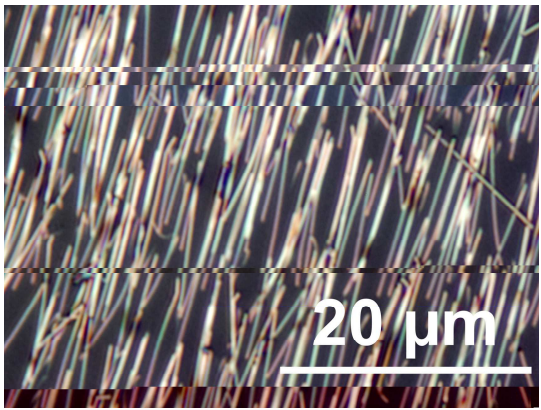}
	\caption{Darkfield optical micrograph of oriented 80 nm diameter germanium nanowires placed flat on a quartz substrate (100X, NA 0.9 objective).  Note: Nanowires appear wider than 80 nm due to diffraction.}
	\label{fig:NW_100}
\end{figure}

\section{Germanium Nanowire Fabrication}
\begin{figure}[tb]
	\centering
		\includegraphics[width=.4\textwidth]{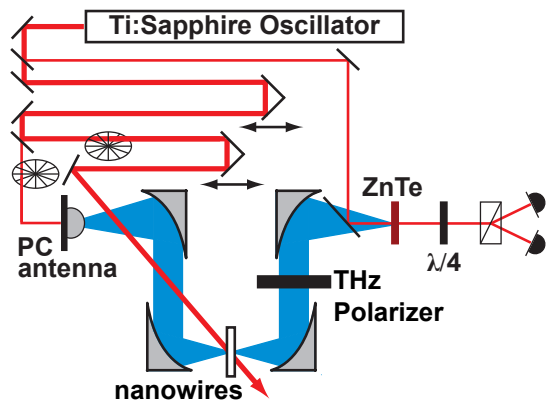}
	\caption{Polarization-sensitive optical-pump THz-probe setup based on a photoconductive THz emitter and electro-optic detection.}
	\label{fig:es}
\end{figure}

Germanium nanowires used in this work were $\sim$80 nm in diameter and $\sim$10 $\mu$m in length (see Fig.~\ref{fig:NW_100}).  They were grown via CVD in a hot-walled quartz tube furnace using germane as the source gas and gold nanoparticles for the catalyst \cite{Cui01}.  Alignment of nanowires was achieved on quartz crystal substrates using a contact printing method previously reported by Fan \emph{et al.}\cite{Fan08}.  Nanowires used in this experiment were unintentionally doped, and expected initial carrier density is less than $10^{17}$ 1/cm$^3$.  Electron-hole pairs were optically generated in the nanowires using 90 fs pulses from a Ti:Sapphire laser with a center wavelength of 780 nm focused to a spot with standard deviation $\sim$150 $\mu$m.  Pump pulse energies in the 1-12 nJ range were used.  The photoexcited nanowires were then probed with a synchronized few-cycle THz pulse generated and detected with a THz time-domain spectrometer (see Fig.~\ref{fig:es}).  The spectrometer, with a power SNR of 4x10$^6$ and measurable frequency range of .5-2.8 THz, was based on a semi-insulating GaAs photoconductive emitter \cite{Katzenellenbogen91} and a ZnTe electro-optic detector \cite{Wu96}.  By varying the delay of the THz pulse with respect to the optical pump pulse, we measured the time-dependent differential change in the THz transmission.  The optical pump and THz probe beams were mechanically chopped at 1400 Hz and 2000 Hz, respectively, and a lock-in amplifier was used to measure the signal at the sum of these frequencies. Measurements in this work were performed at 300K, and the measurement error, due primarily to long-term drift of optomechanical components and the Ti:Sapphire laser, is estimated to be $\sim$5\%.

\begin{figure}[tb]
	\centering
		\includegraphics[width=.47\textwidth]{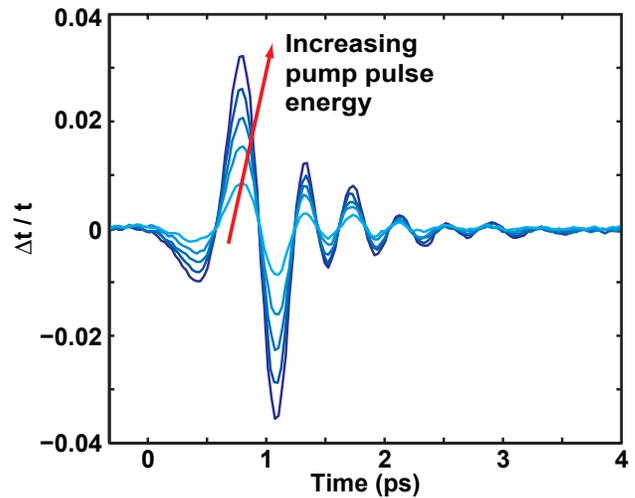}
	\caption{The measured differential amplitudes of THz pulses transmitted through photoexcited germanium nanowires for pump pulse energies of 10.2, 8.2, 6.1, 4.1, and 2.0 nJ are plotted for a fixed pump-probe delay. The THz field is polarized parallel to the nanowires, and the differential amplitudes plotted are scaled by the peak amplitude of the transmitted THz pulse in the absence of photoexcitation.}
	\label{fig:ts}
\end{figure}

\section{Experiments and Results}
Fig.~\ref{fig:ts} shows the measured differential amplitudes of THz pulses transmitted through photoexcited germanium nanowires for pump pulse energies of 10.2, 8.2, 6.1, 4.1, and 2.0 nJ for a fixed pump-probe delay. The THz field is polarized parallel to the nanowires. Since double-chopping is employed, the measured differential signal is affected only by the THz response of the photoexcited carriers within the nanowires. Fig.~\ref{fig:ts} displays no measurable carrier density dependence in the frequency dispersion of the THz response since the measured pulse shape remains unchanged for different pump pulse energies (only the pulse amplitude changes). As discussed in detail below, carrier density independent dispersion is a result of very small plasma frequencies. These results show that the dynamics of photoexcited carriers can be studied by measuring the differential amplitude of the peak of the transmitted THz pulse as a function of the pump-probe delay \cite{Lui01}.

\begin{figure}[tbp]
  \centering
    \includegraphics[width=.47\textwidth]{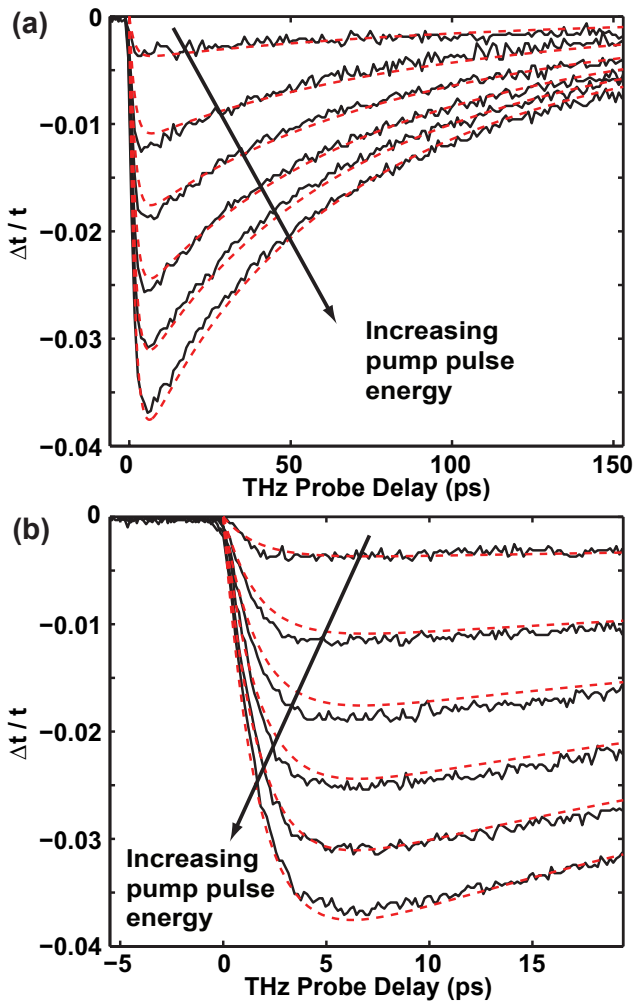}
  \caption{(a) Measured (solid lines) normalized differential amplitude  $\Delta t/t$ of the peak of the THz probe pulse is plotted as a function of the pump-probe delay for optical pump energies of 12 nJ, 9.8 nJ, 7.6 nJ, 5.4 nJ, 3.3 nJ, and 1.1 nJ. The theoretical fit (dashed lines) is also shown. THz probe transmission is seen to recover on a 75-125 ps time scale. (b) Close-up of the differential transmission transient.}
  \label{fig:transients}
\end{figure}

\begin{figure}[tbp]
	\centering
		\includegraphics[width=.47\textwidth]{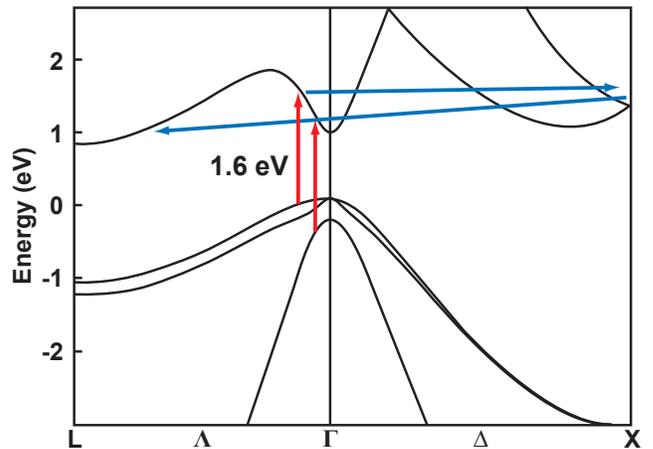}
	\caption{Ultrafast dynamics of photoexcited carriers in a germanium nanowire energy band are depicted \cite{Bailey94}. Electrons are photoexcited near the $\Gamma$ point and quickly scatter to the X point. In the next few picoseconds electrons scatter to the L-valley where they remain until they recombine with the holes.}
	\label{fig:band}
\end{figure}

Fig.~\ref{fig:transients} shows the measured differential amplitude of the peak of the THz probe pulse as a function of the pump-probe delay for different optical pump energies. The THz transmission decreases in the first $\sim$5 ps following the optical excitation and then recovers on a 75-125 ps time scale. These two time scales in the measured transient can be explained by the intraband and interband carrier dynamics, respectively. The optical pulse creates electron-hole pairs near the $\Gamma$-point in the germanium reciprocal lattice (see Fig.~\ref{fig:band}). Electrons quickly scatter from the $\Gamma$-point to the X-point within 100 fs due to strong intervalley phonon scattering, after which they scatter to the lowest L-valley within a few picoseconds \cite{Bailey94}. Photoexcited holes in the three valence bands are also expected to thermalize within 0.5 ps  \cite{Bailey94}. Higher electron mobility in the L-valley increases the THz optical conductivity, so as the L-valley electron density increases, the transmission of the incident THz radiation decreases \cite{Urbanowicz05}. The transmission of the THz radiation recovers as the hole and the L-valley electron densities decrease due to recombination. Using a simple theoretical model discussed below, the fits to the measured transients are also shown in Fig.~\ref{fig:transients}. A characteristic time scale of 1.7 ps is extracted for the initial transmission decrease. This value corresponds to the electron scattering time from the X-point to the L-point and is consistent with the calculated rates and previously reported measurement results \cite{Prasankumar08,Urbanowicz05}. In bulk germanium, carrier recombination rates are highly dependent on the doping density. Recombination times as long as hundreds of microseconds for undoped germanium \cite{Gaubas06} and as short as hundreds of picoseconds for doped germanium \cite{Urbanowicz05} have been reported. Our measurements show that electrons and holes in 80 nm germanium nanowires recombine with carrier density-dependent recombination times between 75-125 ps.  This shorter time scale, compared to that in bulk germanium, indicates that surface defect states may be responsible for faster recombination in agreement with recent electrical and optical pump-probe measurements \cite{Prasankumar08,Leonard09}.

\begin{figure}[tbp]
	\centering
		\includegraphics[width=.47\textwidth]{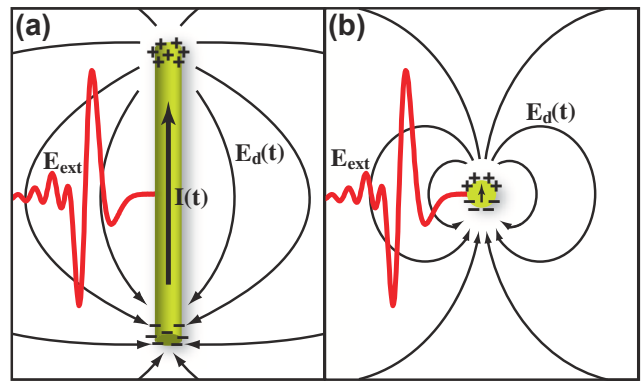}
	\caption{(a) A nanowire oriented parallel to the polarization of the external THz field ($E_{\rm ext}$).  The depolarization field ($E_d$) due to charge confinement on the surface of the nanowire is weak, allowing a Drude-like induced current. (b) A nanowire oriented perpendicular to the external field polarization.  Here, the depolarization field is strong, suppressing the induced current.}
	\label{fig:cartoon}
\end{figure}

\begin{figure}[tbp]
  \centering
    \includegraphics[width=.47\textwidth]{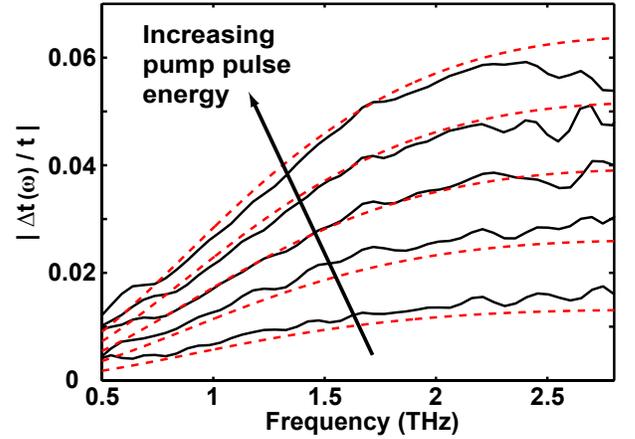}
  \caption{Data (solid) and theory (dashed) for the spectra of the differential THz transmission $|\Delta t(\omega)/t|$ is plotted for different pump pulse energies. The theory assumes a standard deviation of $\sim$150 $\mu$m for the optical intensity profile at the focus and a momentum scattering time $\tau=70\pm15$ fs independent of the carrier density.}
  \label{fig:spectra}
\end{figure}

The THz frequency response of a finite length nanowire can be described by the Drude model modified to consider the depolarization fields \cite{Kittel} due to the induced charges on the surface of the nanowire \cite{Parkinson07,Pitarke07,Nienhuys05}. The inclusion of the depolarization field leads to a surface-plasmon-like resonance in the frequency dependent current response of the nanowire. The current $I(\omega)$ in a nanowire of cross-sectional area $A$ can be written as,
\begin{equation}
I(\omega)=A\frac{\sigma_{\circ}}{1-i\omega \, \tau} \big( E_{\rm ext}(\omega) + E_{d}(\omega) \big)
\end{equation}
where $\sigma_{\circ}$ is the DC conductivity of the nanowire material, $E_{\rm ext}(\omega)$ is the applied field and $E_{d}(\omega)$ is the depolarization field. The above expression can also be expressed as, $I(\omega)=A \sigma_{\rm eff}(\omega) E_{\rm ext}(\omega)$, where the effective conductivity $\sigma_{\rm eff}(\omega)$ is
\begin{equation}
\sigma_{\rm eff}(\omega) = \frac{\sigma_{\circ}}{1-i\omega\,\tau \, (1-\Omega_p^2/\omega^2)}
\label{eq:cond}
\end{equation}
Here, $\tau$ is the momentum scattering time and $\Omega_{p}$ is the frequency of the surface plasmon resonance due to the depolarization field.  $\Omega_p$ is related to the bulk plasma frequency $\omega_{p}$ by a constant factor $g$ that depends on the polarization of the applied field with respect to the nanowire axis (see Fig.~\ref{fig:cartoon}). For a field polarized perpendicular to the nanowire, $g$ equals $\sqrt{\epsilon_{s}/(\epsilon_{s}+\epsilon_{\circ})}$, where $\epsilon_{s}$ and $\epsilon_{\circ}$ are the permittivities of the nanowire material and free-space, respectively. In the case of the field polarization parallel to the nanowire, the value of $g$ is small and is on the order of $(d/L)\sqrt{\epsilon_{s}/\epsilon_{\circ}}$, where $d$ and $L$ are the diameter and length of the nanowire respectively. Since $d\ll L$, when the field is polarized parallel to the nanowires, $\Omega_{p}$ is much smaller than $\omega_{p}$.  We estimate $\Omega_{p}$ to be less than 300 GHz for even the largest photoexcited carrier densities in our experiments. The interaction between nanowires is expected to reduce the depolarization field inside the nanowires and, therefore, further reduce the value of $\Omega_{p}$. At frequencies much larger than $\Omega_p$, $\sigma_{\rm eff}(\omega)$ reduces to the Drude result, and in the DC limit, $\sigma_{\rm eff}(\omega)$ goes to zero as it should for a finite-length uncontacted nanowire. The differential THz transmission (normalized to the transmission in the absence of photoexcitation) can be written as,
\begin{equation}
\frac{\Delta t(\omega)}{t} = \frac{1}{1 + \sigma_{\rm eff}(\omega)\,F(\omega)\,\eta_\circ \, f \, d/(1+n)} - 1 \nonumber
\end{equation}
\begin{equation}
\!\!\!\!\!\quad \approx - \sigma_{\rm eff}(\omega)\,F(\omega)\,\eta_\circ \, f \, d/(1+n)
\label{eq:trans}
\end{equation}
where $\eta_\circ$ is the impedance of free space, $f \approx .08$ is the fill factor of the nanowires, $d = 80$ nm is the diameter of a nanowire, and $n = 1.96$ is the THz refractive index of the quartz substrate \cite{Randall67}. $F(\omega)$ is the overlap factor that accounts for the frequency dependence of the measured THz response due to the mismatch between the optical and THz focus spots.  Assuming Gaussian transverse intensity profiles for the optical and THz beams, the overlap factor is found to be,
\begin{equation}
F(\omega)=\omega^2/(\omega_\circ^2+\omega^2)
\end{equation}
where $\omega_\circ \approx 2\pi c/a$ is approximately the frequency corresponding to the standard deviation, $a=150$ $\mu$m, of the optical beam transverse intensity profile. In the case of the THz field polarized along the nanowires, since $\Omega_{p}\ll\omega$ for frequencies $\omega$ in the 0.5-3.0 THz range, $\sigma_{\rm eff}(\omega)$ has the frequency dependence of the Drude model.  There is therefore no carrier density dependence in the frequency dispersion of the differential THz transmission, in agreement with the measured results shown earlier in Fig.~\ref{fig:ts}. Fig.~\ref{fig:spectra} shows the measured frequency spectra (solid lines) of $|\Delta t(\omega)/t|$ for different pump pulse energies. Also shown are the theoretical fits (dashed lines) obtained from Equation \ref{eq:trans}. As seen in Fig.~\ref{fig:spectra}, the theory agrees well with both the frequency dependence and the carrier density dependence of the data. From our fits, we find the momentum scattering time to be $\tau=70\pm15$ fs, which corresponds to an effective electron plus hole mobility of 4400 cm$^2$/(V$\cdot$sec).  This is slightly smaller than the bulk germanium electron plus hole mobility of 5700 cm$^2$/(V$\cdot$sec) at 300 K found in the literature \cite{Madelung}.

Equation \ref{eq:trans} shows that the differential THz transmission is approximately proportional to the carrier density through $\sigma_{\rm eff}(\omega)$. If the carrier density changes on a time scale much slower than the momentum scattering time, then the differential amplitude of any one point on the transmitted THz pulse measured as a function of the pump-probe delay can be used to study ultrafast carrier dynamics. The time resolution in our experiments is limited by the width of the optical pump pulse to $\sim$150 fs. In order to describe the complete differential THz transmission transient shown in Fig.~\ref{fig:transients}, we model the time dependence of the photoexcited electron density in the germanium nanowires with rate equations. We assume that the photoexcited electron density $N'$ in the higher energy valleys in the conduction band relaxes into the lowest energy L-valley with characteristic time $\tau_{r}$. In the L-valley, the electrons interact with the THz radiation until they recombine \cite{Bailey94}. Recombination in bulk germanium with low doping is generally attributed to the Shockley-Reed-Hall (SRH) mechanism of defect assisted recombination. Auger recombination becomes dominant for doping densities above $10^{18}$ cm$^{-3}$ \cite{Gaubas06}. Surface defect recombination in nanowires is also expected to have carrier density dependence similar to that of the bulk SRH mechanism. We assume that the recombination rate is described by a second-order polynomial in the L-valley electron density $N$,
\begin{equation}
\frac{dN'}{dt} = -\frac{N'}{\tau_{r}} \qquad\qquad \frac{dN}{dt} = \frac{N'}{\tau_{r}}-(A\ N+B\ N^2)
\end{equation}
The initial photoexcited density $N'(t=0)$ is estimated to be $4.5\times 10^{18}$ 1/cm$^3$ for a 12 nJ pump pulse and is assumed to scale linearly with the pump pulse energy. The DC conductivity $\sigma_{\circ}$ equals $Ne(\mu_{e} + \mu_{h})$, where $\mu_{e}$ and $\mu_{h}$ are the electron and hole mobilities. The agreement between the rate equation model and the data is shown in Fig.~\ref{fig:transients}. The extracted values of the various parameters for best fit are as follows: $\tau_{r}=1.7$ ps, $A = 8.8\times10^{9}$ 1/sec, and $B = 2\times10^{-9}$ cm$^3$/sec. The model agrees with the data for all pulse energies. The necessity of the $B$ parameter indicates carrier density-dependent recombination rates in germanium nanowires. This is consistent with density-dependent SRH surface and Auger recombination \cite{Muller03}.

\begin{figure}[tb]
  \centering
    \includegraphics[width=.30\textwidth]{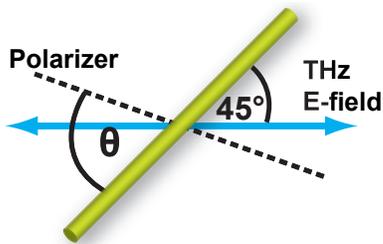}
  \caption{The incident THz polarization is fixed at 45$^\circ$ with respect to the nanowire axis. A polarizer is rotated at an angle $\theta$ with respect to the nanowire axis to select THz polarization post-transmission.}
  \label{fig:wire-pol}
\end{figure}

\begin{figure}[tb]
  \centering
    \includegraphics[width=.47\textwidth]{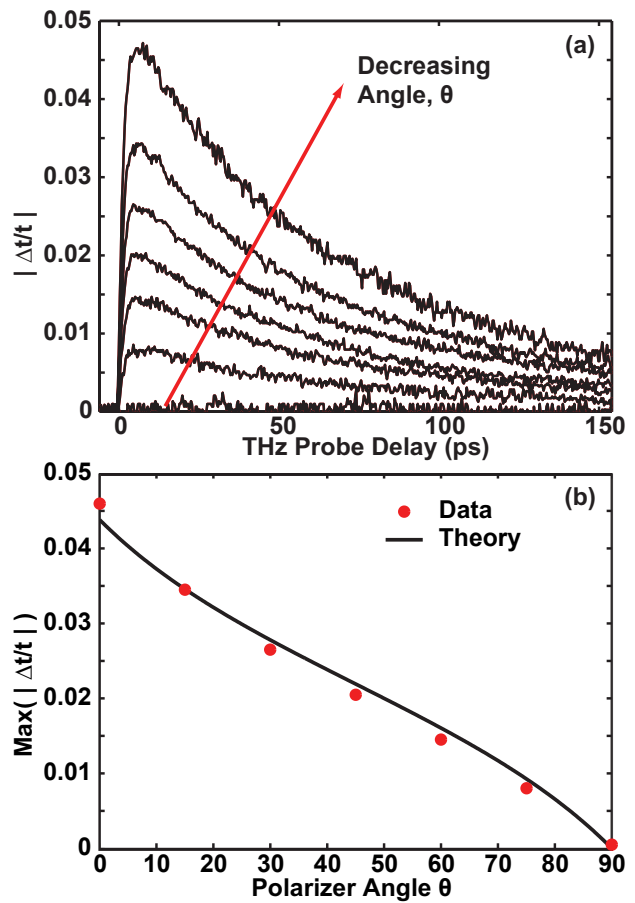}
  \caption{(a) Measured differential terahertz transmission $|\Delta t/t|$ for polarizer angles of $\theta=0, 15, 30, 45, 60, 75, 90$ degrees.  At $\theta = 90$ degrees, $|\Delta t/t|$ is negligibly small and lost in the noise.  (b) Maximum values of $|\Delta t/t|$ from Fig.~\ref{fig:transpol}(a) are plotted versus the polarizer angle $\theta$. The angular dependence expected from the theory is also plotted (solid line).}
  \label{fig:transpol}
\end{figure}

Photoexcited carriers in oriented nanowires are expected to exhibit a polarization dependent THz response due to the geometries depicted in Fig.~\ref{fig:cartoon}. In order to study the polarization dependence of the THz transmission, the incident THz electric field was polarized at 45$^\circ$ with respect to the nanowires (see Fig.~\ref{fig:wire-pol}). As a result, the field had components both parallel and perpendicular to the nanowires. After transmission through the nanowire sample, the field polarization was selected for measurements of $\Delta t/t$ by rotating the polarizer through an angle $\theta$ with respect to the nanowires. Fig.~\ref{fig:transpol} shows the measured values of $|\Delta t/t|$ for different angles $\theta$. The most striking feature of the data is the absence of any measurable THz response when the field is polarized perpendicular to the nanowires. In this case, the plasma frequency $\Omega_{p}$ equals $\omega_{p} \sqrt{\epsilon_{s}/(\epsilon_{s}+\epsilon_{\circ})}$ and is in the tens of THz range for the photoexcited carrier densities in our experiments. Equation \ref{eq:cond} shows that when $\Omega_{p}\gg\omega$, and the product $\omega \, \tau$ is not too small, the induced current is significantly reduced compared to the case when $\Omega_{p}\ll\omega$. Therefore, for perpendicular THz field components the depolarization field is strong enough to suppress the induced current, and so the resulting THz response is much weaker compared to that for parallel components. In this way, the shape anisotropy of the nanowires on subwavelength scales determines the polarization dependence of the THz response. Assuming that the THz response of oriented nanowires is negligibly small when the field is polarized perpendicular to the nanowires, the measured THz transmission is expected to be proportional to the cosine of the angle between the field polarization and the nanowire axis. In our experiments, since the field polarization is selected post-transmission, the measured values of $|\Delta t/t|$ are expected to be proportional to $\cos(\theta)/\cos(\pi/4-\theta)$. Fig.~\ref{fig:transpol}(b) shows that the (peak) values of $|\Delta t/t|$ exhibit exactly this angular dependence. 

\section{Conclusion}
In conclusion, we have measured the time, frequency, and polarization dependence of the THz response of germanium nanowires using optical-pump terahertz-probe spectroscopy. Carrier intraband relaxation times, interband electron-hole recombination times, and carrier momentum scattering times were also measured using the same technique. Our results demonstrate the usefulness of ultrafast THz spectroscopy for characterizing nanostructured materials. 

\acknowledgements
The authors acknowledge support from Eric Johnson through the National Science Foundation, the DARPA Young Faculty Award under Ronald Esman, and the National Science Foundation through the CAREER award.  The authors also acknowledge the DARPA TBN program and the Cornell CCMR REU program.  

\bibliography{optpnw}

\end{document}